\documentclass[twocolumn,prb,showpacs]{revtex4}
\usepackage{graphics}
\begin{document}
\title{Ground state of the random-bond spin-1 Heisenberg chain}
\author{Sara Bergkvist}
\email{sara@theophys.kth.se}
\author{Patrik Henelius}
\author{Anders Rosengren}
\affiliation{Condensed Matter Theory,Physics Department, KTH,
             SE-106 91 Stockholm, Sweden}
\date{\today}

\begin{abstract}
Stochastic series expansion quantum Monte Carlo is used to study the
ground state of the antiferromagnetic spin-1 Heisenberg chain with
bond disorder. Typical spin- and string-correlations functions behave
in accordance with real-space renormalization group predictions for
the random-singlet phase. The average string-correlation function
decays algebraically with an exponent of -0.378(6), in very good
agreement with the prediction of $-(3-\sqrt{5})/2\simeq -0.382$, while
the average spin-correlation function is found to decay with an
exponent of about -1, quite different from the expected value of
-2. By implementing the concept of directed loops for the spin-1 chain
we show that autocorrelation times can be reduced by up to two orders
of magnitude.

\end{abstract}
\pacs{75.10.Jm, 75.40.Mg, 75.50.Ee }
\maketitle

\section{Introduction} 
The ground state of quantum spin chains displays a rich variety of
physical phenomena such as quasi-long-range order, quantum phase
transitions and hidden order parameters. Some models are exactly
solvable by Bethe ansatz techniques, while others have to be
approached using approximate techniques such as spin-wave theory,
renormalization group (RG) methods, bosonization or numerical
simulations.\cite{review} Some methods, such as the density matrix
renormalization group technique,\cite{WhPRL92} have been developed in
the context of spin chains.

The introduction of disorder into quantum spin chains appears to
create a very complex problem, yet there has been remarkable success
in understanding and solving certain strongly disordered spin
chains. Using a real-space renormalization group method Ma, Dasgupta
and Hu were able to obtain many results for the bond-disordered
spin-1/2 Heisenberg chain.\cite{MaPRL79} Fisher managed to solve the
RG equations exactly for the spin-1/2 case,\cite{FiPRB94} and he
showed that the ground state is the so-called random-singlet phase,
where all spins pair up and form singlets over arbitrarily large
distances. The characteristic time scale is an exponential function of
the length scale, resulting in a dynamic critical exponent
$z=\infty$. The random-singlet phase is gapless and average
correlation functions decay algebraically, while typical correlations
decay exponentially. Due to the singlet coupling between spins the
decay exponents are predicted to be equal for all components of the
correlation function, even if the Hamiltonian is anisotropic. In the
RG calculation any initial randomness flows to the infinite-disorder
random-singlet fixed point. Many of these striking prediction have
been confirmed by numerical
studies.\cite{YoPRB96,HiJPSJ96,JoPRB97,HePRB98}

The physics of the clean spin-1 chain is dramatically different from
the spin-1/2 chain due to the presence of the Haldane
gap,\cite{HaPRL83} accompanied by exponentially decreasing
spin-correlation functions and a non-zero string-order parameter. The
effects of adding disorder to the spin-1 chain are more controversial,
since the RG equations cannot be solved analytically in this case. By
mapping the spin-1 chain to an effective spin-1/2 system Hyman and
Yang argue that the spin-1 chain is stable against weak
disorder,\cite{HyPRL97} but for a sufficiently strong disorder the
chain undergoes a second order phase transition to the random-singlet
phase. A density matrix renormalization group study finds no evidence
of such a transition,\cite{HiPRL99} but Hyman and Yang suggest that the
disorder was not strong enough in the numerical study.\cite{HyPRL00}
However, for the same disorder distribution a recent quantum Monte
Carlo (QMC) study finds evidence of a transition to the
random-singlet phase,\cite{ToJPSJ00} and a new RG study also supports
this result.\cite{Sa02}

In a further attempt to resolve the nature of the ground state of the
strongly disordered spin-1 chain our goal is to study the average and
typical spin- and string-correlation functions. The decay and
distribution of these functions is one of the hallmarks of the
random-singlet phase, and has been studied numerically to confirm the
random-singlet picture for the spin-1/2 chain.\cite{YoPRB96,JoPRB97,HePRB98}
To obtain this goal we perform QMC simulations at a temperature low
enough so that the observables in question have obtained their ground
state expectation values. We use the stochastic series expansion QMC
algorithm\cite{SaPRB99} in our calculations, and by applying the
recently introduced concept of directed loops~\cite{Sy02} to the
spin-1 chain we demonstrate that autocorrelation times can be
decreased by up to two orders of magnitude in the cases we have
examined.

The outline of the paper is as follows: in Sec.~\ref{sec2} we
review the basics of the stochastic series expansion. In
Sec.~\ref{sec3} we implement the method of directed loops for the
spin-1 case, and show the effects on the autocorrelation times.  The
method used to determine average and typical correlations is discussed
in Sec.~\ref{corr}. Results for the ground state behavior of the
random-bond spin-1 chain are shown in Sec.~\ref{sec4}. We conclude
with summary and discussion in Sec.~\ref{sec5}.
 
\section{Operator-loop algorithm}
\label{sec2}
The stochastic series expansion is described in detail
elsewhere.\cite{SaPRB99} In this section we give a brief summary in
order to explain the new features we introduce in Sec.~\ref{sec3}.

The model treated in this article is the one-dimensional
antiferromagnet spin-1 Heisenberg  chain with bond disorder. The
Hamiltonian, $H$, is given by,
\begin{equation}
H=\sum_{i}J_{i}{\bf S}_i\cdot {\bf S}_{i+1},
\end{equation}
where ${\bf S}$ denotes a spin-1 operator and the positive coupling
parameter $J_{i}$, is randomly distributed.  

The partition function, $Z$, is Taylor
expanded,
\begin{equation}
Z=\sum_{\alpha}\sum_{m=0}^{\infty}\frac{(-\beta)^m}{m!}
\langle\alpha|H^m|\alpha\rangle,
\label{Z}
\end{equation}
in the basis $\{|\alpha\rangle\}=\{|S_1^z,S_2^z,..,S_N^z\rangle\}$,
where $N$ is the number of spins. The inverse temperature is denoted by
$\beta$.

Next we rewrite the Hamiltonian as a sum over diagonal and
off-diagonal operators,
\begin{equation}
\label{ham}
H=-\sum_{b=1}^{N}J_{b}(H_{1,b}-H_{2,b}),
\end{equation}
where $N$ is the number of spins and $b$ denotes a bond corresponding
to a pair of interacting spins $j(b)$ and $k(b)$. The operators are
given by
\begin{eqnarray}
\label{energies}
H_{1,b}&=&C-S^z_{j(b)}S^z_{k(b)}\\
H_{2,b}&=&\frac12(S^+_{j(b)}S^-_{k(b)}+S^-_{j(b)}S^+_{k(b)}),\nonumber
\end{eqnarray}
where $C$ is a constant inserted to assure that the weight function is
positive for all configurations. 

To simplify the Monte Carlo update we introduce an additional unit
operator $H_{0,0}=1$.  Inserting the Hamiltonian given by
Eq.~(\ref{ham}) into Eq.~(\ref{Z}), and truncating the sum at $m=L$ we
obtain
\begin{equation}
Z=\sum_{\alpha}\sum_{S_L}\frac{\beta^n(L-n)!}{L!}\langle
\alpha|\prod_{i=1}^L J_{b_i}H_{a_i,b_i}|\alpha\rangle,
\label{Zeq}
\end{equation}
where $S_L$ denotes a sequence of operator-indices
\begin{equation}
S_L=(a_1,b_1)_1,(a_2,b_2)_2,...,(a_L,b_L)_L,
\end{equation}
with $a_i={1,2}$ and $b_i={1,\ldots,N}$ or $(a_i,b_i)=(0,0)$ and $n$ is
the number of non-unit operators in $S_L$.

The Monte Carlo procedure must hence sample the space of all states
$|\alpha\rangle$, and all sequences $S_L$. The simulation starts with
some random state $|\alpha\rangle$ and an operator string containing
only unit operators. One Monte Carlo step consists of a diagonal and
an off-diagonal update.  In the diagonal update attempts are made to
exchange unit and diagonal operators sequentially at each position in
the operator string. Defining the propagated state
\begin{equation}
|\alpha(p)\rangle\sim\prod_{i=1}^{p}H_{a_i,b_i}|\alpha\rangle,
\end{equation}
the probability for inserting or deleting a diagonal operator at place
$p$ in the operator string is given by detailed balance,\cite{SaPRB99}
\begin{eqnarray}
P([0,0]_p\rightarrow[1,b]_p)=\frac{ J_b N\beta\langle
\alpha(p-1)|H_{1,b}|\alpha(p-1)\rangle}{L-n}\\
P([1,b]_p\rightarrow[0,0]_p)=\frac{L-n+1}{J_b N\beta\langle
\alpha(p-1)|H_{1,b}|\alpha(p-1)\rangle }\nonumber.
\end{eqnarray}
The inclusion of $J_b$ in the expressions above represents the place
where the bond disorder appears explicitly in the algorithm. Removing
or inserting an operator changes the expansion power $n$ by $\pm 1$.

The off-diagonal update, also called loop update, is carried out with
$n$ fixed.  Each bond operator $H_{b_i}=H_{1,b_i}+H_{2,b_i}$ acts only
on two spins, $S_{j(b_i)}$ and $S_{k(b_i)}$. We can therefore rewrite
the matrix elements in Eq.~(\ref{Zeq}) as a product of $n$ terms,
called vertices,
\begin{equation}
M(\alpha,S_L)=\prod_{i=1}^{n}W_{v(i)},
\label{vertw}
\end{equation}
where the vertex weight $W_{v(i)}$ is defined as
\begin{eqnarray}
&&W_{v(i)}= \nonumber\\ 
&&\langle S^z_{j(b)}(p)S^z_{k(b)}(p)|H_{b_i}|S^z_{j(b)}(p-1)S^z_{k(b)}(p-1)
\rangle .\hspace{10pt}
\label{vweight}
\end{eqnarray}
A vertex thus consists of four spins, called the legs of the vertex,
and an operator.

The principles of the loop update are: one of the $n$ vertices is
chosen at random and one of its four legs is randomly selected as the
entrance leg. The entrance leg is given a random state that differs
from its initial state. One of its four legs is chosen as the exit
leg, and its state is also changed. In order to satisfy detailed
balance the exit leg is chosen with a probability proportional to the
vertex weight $W_{v(i)}$ after the spins have been flipped. Thereafter
the vertex list is sequentially searched for the next vertex that
contains the exit spin.  This spin becomes the entrance leg of the
next vertex and this procedure is continued until the loop reaches the
original entrance leg, when the loop closes. During one Monte Carlo
step the loop update is repeated until on average all the vertices
have been updated.

\section{Directed loops for spin-1 chain}
\label{sec3}

The operator-loop formulation described above is applicable to a wide
variety of models.\cite{He02} The efficiency of the algorithm does,
however, depend on the model in question. It is particularly efficient
for cases where the so-called bounce or backtrack process can be
avoided.  If the entrance and exit legs of a vertex coincide, the
vertex spin configuration is left unchanged and the loop backtracks to
the previous vertex. This is undesirable since the bounce process does
not achieve any change in the SSE configuration. For models such as
the spin-1/2 Heisenberg model and spin-1/2 XX model, where the bounce
process can be avoided, the autocorrelation times are generally
significantly shorter than for cases where the bounce has to be
included.

Recently Sandvik and Sylju{\aa}sen showed that, by introducing the
notion of directed loops, it is possible to exclude the bounce process
for a spin-1/2 chain in a wide parameter regime.\cite{Sy02} The
application to more general models is also discussed in
Ref.~\onlinecite{Sy02}, as well as in Ref.~\onlinecite{Ha02}.  We will
now show in some detail how the bounce process can be eliminated also
for the spin-1 case.

The detailed balance principle requires that  the probabilities of
changing between two state, $s$ and $s'$, with weight functions $W(s)$
and $W(s')$, satisfy
\begin{equation}
W(s)P(s\rightarrow s')=W(s')P(s'\rightarrow s).
\end{equation}
By considering all possible loop configurations that take us from
configuration $s$ to $s'$, as well as the ``time''-reversed paths
returning to configuration $s$, Sandvik and Sylju{\aa}sen showed that
the detailed balance criteria is satisfied if the local vertex
condition
\begin{equation}
W_sP(s,l_1\rightarrow s',l_2)=W_{s'}P(s',l_2\rightarrow s,l_1)
\label{det_bal1}
\end{equation}
is fulfilled. $W_s$ denotes the vertex weight, given by
Eq.~(\ref{vweight}), and $P(s,l_1\rightarrow s',l_2)$ is the probability
to choose exit leg $l_2$, given entrance leg $l_1$. The spin
configuration of the vertex changes from $s$ to $s'$. This criteria
relates the transfer probabilities of two processes where the
direction of the loop segment is reversed, since the entrance and exit
legs are interchanged. Considering that the probabilities for
choosing different exit legs should sum to unity Eq.~(\ref{det_bal1})
immediately leads to
\begin{equation}
W_s= \sum_{l_2}W_sP(s,l_1\rightarrow s',l_2).
\label{det_bal2}
\end{equation}
We can use Eq.~(\ref{det_bal1}) and Eq.~(\ref{det_bal2}) to construct
the probabilities, $P(s,l_1\rightarrow s',l_2)$, needed to perform the
loop update. It turns out that it is possible to divide the space of
all processes into subspaces, where only vertices in the same group
are related to each other by Eq.~(\ref{det_bal1}). Hence, one may
solve the detailed balance equations for each transformation group
separately, and in many cases it is possible to find a solution with
no bounce processes.

\begin{figure}
\begin{center}
\resizebox{60mm}{!}{\includegraphics{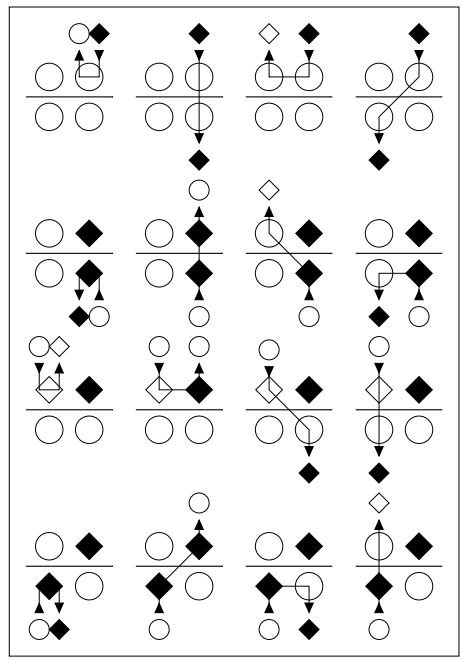}}
\end{center}

\caption{An example of a transformation group with four vertices. The
solid (empty) diamonds represent a spin projection of +1 (-1), and the
circles the projection of 0. The lines with arrows are the directed
loop segments. A row in the figure shows all possible updates of the
vertex for a given entrance leg and spin state.  The smaller symbols
at the beginning and end of the loop segment show the entrace and exit
spin states after the update.}
\label{4group}
\end{figure}

We now show how to apply this method to the spin-1 chain, and how to
solve the resulting equations in a manner that eliminates the bounce
process also in this case.  For the spin-1 model there are in total 52
transformation groups, but many of them can be transformed into each
other by symmetry arguments. There are five irreducible groups with
two vertices, two groups with three elements, and one group with four
vertices. The last group is shown in Fig.~\ref{4group}, where the
operator of each vertex is shown as a horizontal line. The operator
acts on the two spins above it, resulting in the spin configuration
shown below it. Each row represent all vertices that can be reached
from a given entrance leg with a given spin projection.
Eq.~(\ref{det_bal2}) gives us the following set of equations
\begin{eqnarray}
W_1&=&a_1+b+c+d\\
W_2&=&a_2+b+e+f\nonumber\\
W_3&=&a_3+c+e+g\nonumber\\
W_4&=&a_4+d+f+g.\nonumber
\end{eqnarray}
The weight on the left-hand side refers to the weight of the bare
vertex, given by Eq.~(\ref{vweight}), while the weights on the
right-hand side refer to the product of vertex weight and the transfer
probability. Processes that are related to each other by changing the
direction of the loop segment, and interchanging spin configurations,
are given equal weights, according to Eq.~(\ref{det_bal1}).  It is
possible to solve this set of equations under the restriction that all
bounce processes $a_i$ are zero. There is not a unique solution since
the number of unknowns exceeds the number of equations, and we have
chosen the solution which gives all the allowed vertices equal weight.

\begin{figure}
\begin{center}
\resizebox{65mm}{!}{\includegraphics{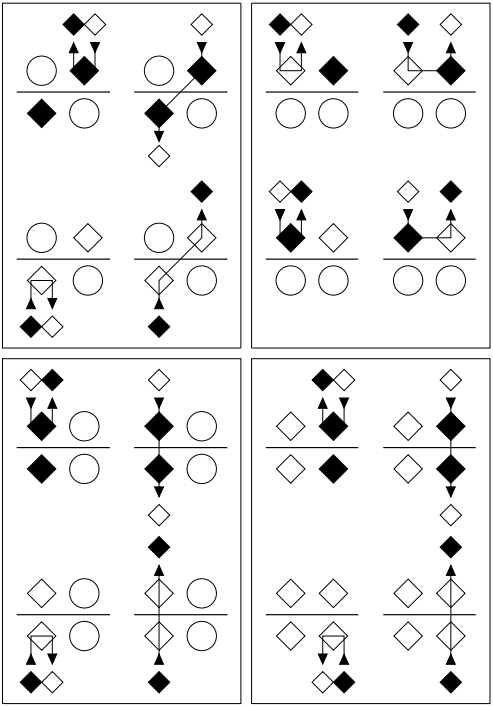}}
\end{center}
\caption{The four transformation groups for the spin-1 chain, in which
the magnetization changes by two. The solid (empty) diamonds represent
a spin projection of +1 (-1), and the circles a projection of 0.}
\label{2change}
\end{figure}

For the spin-1 case the groups can be classified according to the
change in spin projection on the entrance spin. All the groups where
the spin projection changes by one can be solved so that the bounce is
eliminated. There are, however, also four groups where the spin
projection changes by two, shown in Fig.~\ref{2change}. The bounce
cannot be eliminated in this case. The equations for the group in the
bottom right corner are of the form
\begin{eqnarray}
W_1&=&a_1+b=C+1\\
W_2&=&a_2+b=C-1,\nonumber
\end{eqnarray}
and the bounce processes $a_1$ and $a_2$ cannot be eliminated. We can,
however, solve all groups shown in Fig.~\ref{2change} so that the
\textit{only} processes allowed are the bounces. This means that any
attempt to change the magnetization by two will lead to an immediate
bounce and the loop will close. In practice this idea can be
implemented by attempting to change the magnetization only by one, and
accepting 50\% of the attempts when the initial spin state is $\pm 1$.

\begin{figure}
\begin{center}
\resizebox{!}{60mm}{\includegraphics{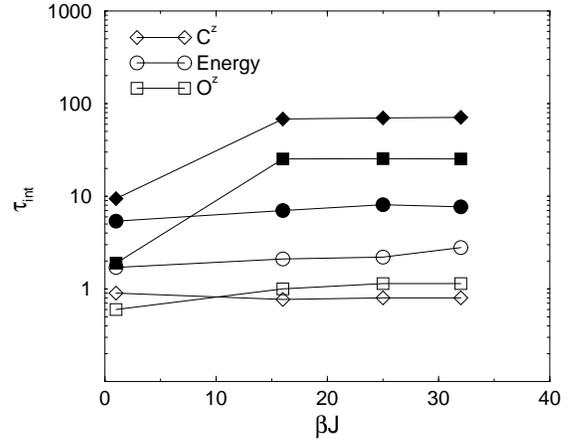}}
\end{center}
\caption{Integrated autocorrelation times as a function of inverse
temperature, $\beta J$, for a system with 32 spins without bond
disorder. The open symbols represents the directed loop updating
method, where bounces have been excluded and the solid symbols are
calculated using the original loop move.}
\label{AuC}
\end{figure}

To study the efficiency of the new loop move, integrated
autocorrelation times have been calculated. The normalized
autocorrelation function is defined as,
\begin{equation}
A_Q(t)=\frac{\langle Q(i+t)Q(i)\rangle-\langle Q(i)\rangle^2}{\langle
Q(i)^2\rangle-\langle Q(i)\rangle^2},
\end{equation}
where $i$ and $t$ are measured in units of Monte Carlo steps, with one
Monte Carlo step defined as one diagonal update and one loop
update. The integrated autocorrelation time is defined according
to
\begin{equation}
\tau_{\mathrm{int}}[Q]=\frac{1}{2}+\sum_{t=1}^{\infty}A_Q(t).
\end{equation}
In Fig.~\ref{AuC} integrated autocorrelation times are shown for
calculations using the old and the new loop updating method. The
calculations are done on a system consisting of 32 spins without any
disorder in the coupling parameters. The autocorrelation times are
calculated for the energy and for the spin- and string-correlation
function, which we now define.

The spin-correlation function $C^z(N)$ for a chain with $N$ spins is
measured at the greatest distance between two spins, which for periodic
boundary conditions equals $N/2$,
\begin{equation}
C^z(N)=\frac{1}{N}\sum_{i=1}^N \langle S^z_iS^z_{i+\frac{N}{2}}\rangle,
\label{spinc}
\end{equation}
and the string-correlation function $O^z(N)$  is similarly defined by,
\begin{eqnarray}
O^z(N)=&&\frac{1}{N}\sum_{i=1}^N \langle S^z_i
\exp[i\pi(S^z_{i+1}+S^z_{i+2}+\cdots \nonumber \\
&&+S^z_{i+\frac{N}{2}-1})]
S^z_{i+\frac{N}{2}}\rangle.
\label{strc}
\end{eqnarray}
The energy is calculated according to\cite{SaJPA92}
\begin{equation}
E=-\frac{1}{\beta}\langle n\rangle,
\label{EnergyOp}
\end{equation}
where $n$ is the number of non-unit operators in the operator string.
The energy autocorrelation time is least affected by the new loop
update, as is clearly seen in Fig.~\ref{AuC}. The reason is that the
energy is not directly dependent on the loop move, since the energy is
given by the number of operators which is determined in the diagonal
update. The exclusion of the bounce shortens the autocorrelation time
by up to two orders of magnitude for the spin correlation at low
temperatures.

\section{Determining average and typical correlations}
\label{corr}

To calculate ground state expectation values in strongly disordered
systems using QMC is not an easy task. The standard deviation in the
disorder averaged expectation values has two sources: the standard
statistical fluctuations, and the sample to sample variation.  In
strongly disordered systems the latter is often dominant and it is
therefore advantageous to perform short simulations of a large number
of disorder configurations. This is problematic for two
reasons. First, each sample requires an adequate equilibration time,
which becomes a very significant part of the simulation, particularly
for short simulations. Second, it may be hard to determine at what
temperature the expectation values have converged to their ground
state values. Repeating the calculations at many temperatures is
very time demanding.
 
Recently, a technique was proposed to alleviate these
difficulties.\cite{Sa01} The basic idea is to perform a very small
number of Monte Carlo steps at a series of decreasing
temperatures. The method builds on two realizations. First, given an
equilibrium Monte Carlo configuration at inverse temperature $\beta$
it is possible to construct an approximate equilibrium configuration
at inverse temperature $2\beta$. Second, due to the extremely short
autocorrelation times equilibrium is reached quickly. Due to the small
number of Monte Carlo steps necessary at each temperature the methods
allows very rapid calculation of disorder averages down to very low
temperatures.

Now we will briefly describe the implementation of the method.  The
energy expectation value, see Eq.~(\ref{EnergyOp}), is proportional to
the length of the operator string and the temperature.  This means
that the length of the operator string is proportional to the inverse
temperature $\beta$ as the ground state is approached.  The
calculation begins at a high temperature, where a few equilibration
Monte Carlo steps are performed, followed by a few steps when
measurements are carried out. Thereafter the temperature is halved at
the same time as the operator string is doubled.  The procedure is
repeated and results in a series of measurements at decreasing
temperatures. This is carried on until disorder averages of the
expectation values show no further temperature dependence.

In the random-singlet phase the spin-correlation functions have a very
broad distribution. The average of the correlation functions will be
dominated by very strong correlations, while the typical correlations
are much weaker. In order to characterize this behavior we measure not
only the average of the correlation function, but also the average of
the logarithm of the correlation function, which defines the typical
correlation function,
\begin{equation}
\log C^z_{\text{typ}}(N)=\frac{1}{N}\sum_{i=1}^N 
\log\langle S^z_iS^z_{i+\frac{N}{2}}\rangle.
\end{equation}
The method described above is suitable for determining average
correlation functions. However, it cannot be used to accurately
determine typical correlation functions. The reason is that small
values of the correlation functions give large contributions to the
typical correlations. Therefore, the whole distribution of the
correlation function needs to be determined quite accurately to
investigate the typical behavior. In the above method the statistical
errors for a single disorder configuration are very large, due to the
small number of Monte Carlo steps. Hence the weak correlations, which
are important for the typical correlations, drown in the statistical
noise. In particular, many negative values will have to be discarded
since the logarithm is defined only for positive arguments.  In order
to determine typical correlation functions we have therefore done the
simulation in two stages. First we use the above method to determine
the temperature at which average correlation functions have converged
in temperature. Thereafter we perform much longer simulations at this
temperature to determine the typical correlation functions.

\section{Results for the random-bond chain}
\label{sec4}

The properties of the random-singlet phase are predicted to be
independent of the underlying Hamiltonian, as long as the ground state
of two nearest-neighbor spins is a singlet. In particular, the
decay exponents of the spin and string-correlation functions should be
the same for the spin-1 and spin-1/2 chain.\cite{HyPRL97} In the
random-singlet phase it is predicted that the average correlation
functions decay algebraically,\cite{FiPRB94}
\begin{equation}
C^z(N)\propto N^{-2}
\end{equation}
for large $N$, and 
\begin{equation}
O^z(N)\propto N^{-\frac{3-\sqrt{5}}{2}}
\simeq N^{-0.382},
\end{equation}
while typical correlations
decay with a stretched exponential as
\begin{equation}
C^z_{typ}(N)\propto \exp(-A\sqrt{N}),
\label{typdec}
\end{equation}
where $A$ is a non-universal constant.

\begin{figure}
\begin{center}
\resizebox{!}{60mm}{\includegraphics{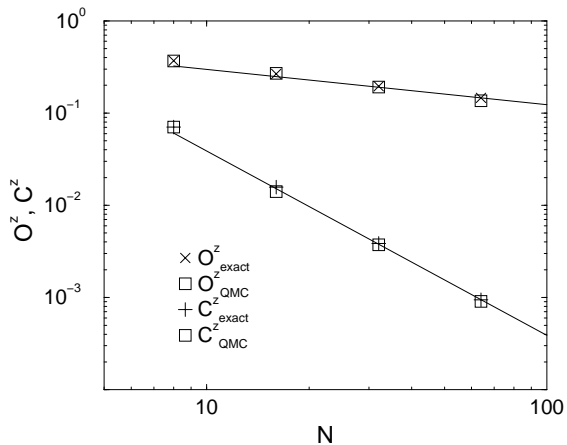}}
\end{center}
\caption{A comparison between exact diagonalization and QMC results
for the correlation functions of the spin-1/2 XX chain with bond
disorder. Lines show algebraic decay with exponents predicted from RG
calculations. The lines are drawn through the last exact
diagonalization data point.}
\label{XX}
\end{figure}

To make sure that the QMC method is capable of correctly determining
the correlation functions for a strongly disordered spin system we
have first performed a calculation on the spin-1/2 XX chain with bond
disorder. The XX chain is described by the following Hamiltonian,
\begin{equation}
H=\sum_{i}J_{i}(S_i^xS_{i+1}^x+S_i^yS_{i+1}^y),
\end{equation}
where ${\bf S}$ denotes a spin-1/2 operator. The ground state
expectation values of the correlation functions for this system have
been calculated by exact diagonalization using the Jordan-Wigner
transformation to free fermions.\cite{HePRB98} We compare QMC results
for the average correlation functions with the diagonalization data in
Fig.~\ref{XX} and there is very good agreement. The lines are drawn
through the last diagonalization data point with the predicted slopes
of -2 for the spin-correlation function and -0.382 for the
string-correlation function. We can see that already for these
relatively small system sizes the data displays almost linear
behavior with slopes close to the predicted values. The finite-size
corrections are more pronounced for the string-correlation function.

\begin{figure}
\begin{center}
\resizebox{!}{70mm}{\includegraphics{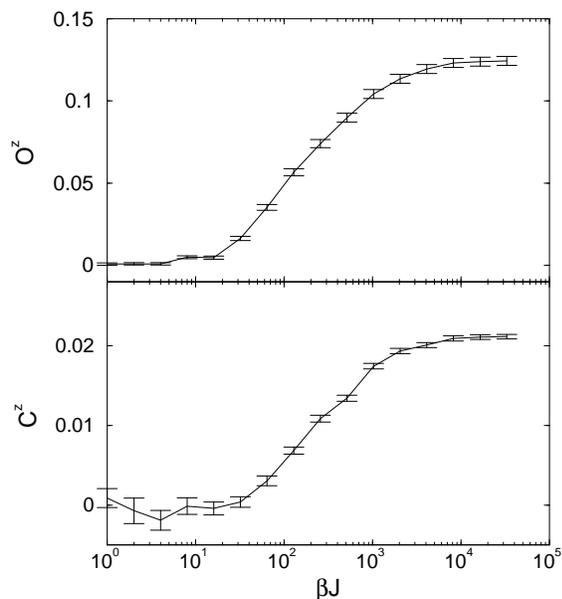}}
\end{center}
\caption{The temperature dependence of the string- and spin-correlation
function for a system with 64 spins.}
\label{TdepStrCorr}
\end{figure}

Next we consider the spin-1 chain. Recent numerical studies of the
disordered spin-1 chain have used bonds distributed according to a box
distribution
\begin{equation}
P(J)=\left\{\begin{array}{cl}
1/W & \mathrm{for~~} 1-\frac{W}{2}\leq J_i\leq 1+\frac{W}{2}\\
0 & \mathrm{otherwise.} \end{array}\right.
\end{equation}
A density matrix renormalization group study did not find a transition
to the random-singlet phase,\cite{HiPRL99} even for the strongest
disorder $(W=2)$, while a QMC study\cite{ToJPSJ00} found evidence for
a transition around $W=1.8-1.9$. According to the RG analysis by Hyman
and Yang\cite{HyPRL00} the phase transition occurs for the power law
distribution, $P(J)\propto J^{-0.33}$, which represents stronger
disorder than the box distribution. Finally, a recent RG
study\cite{Sa02} finds that the transition occurs exactly at $W=2$.
Close to the critical point the RG studies find that the Haldane phase
is Griffiths-like, with no gap, finite string order and exponentially
decaying spin-correlation functions.

In this work we do not attempt to find the critical bond distribution.
Instead we try to resolve the controversy concerning the box
distribution.  We focus on the widest possible box distribution and
make an extensive QMC simulation of the correlation functions to
determine whether they behave according to the random-singlet
predictions.

The temperature dependence of the average string- and spin-correlation
function for system size $N=64$ is shown in Fig.~\ref{TdepStrCorr}.
At each temperature we performed 20 equilibration steps followed by 60
measurement steps. The results seem to have converged (within
statistical error bars) at approximately $\beta\approx32000$.  The
convergence temperatures for different system sizes are presented in
Table~\ref{tab2}. At least 1000 disorder distributions are used in the
calculations. For these system sizes the convergence temperature is
given by $\beta=N^3/8$.

\begin{figure}
\begin{center}
\resizebox{!}{60mm}{\includegraphics{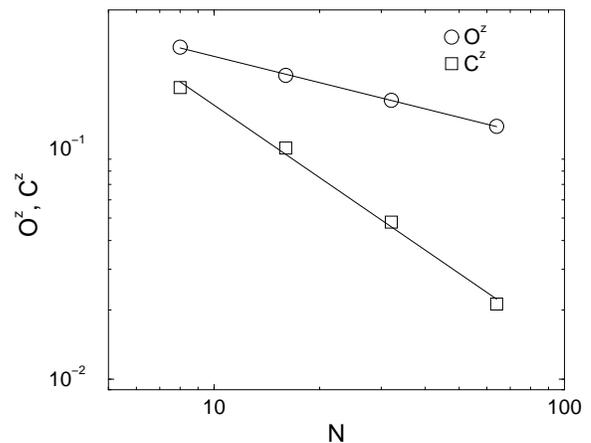}}
\end{center}
\caption{The spin- and string-correlation functions for the
bond-disordered spin-1 chain. Disorder averages are taken over at least
1000 configurations. The solid lines are linear fits to the data
points.}
\label{corrstr}
\end{figure}

\begin{table}[b]
\begin{ruledtabular}
\begin{tabular}{c|c}
 System size & Inverse temperatures $\beta J$\\\hline 8 & 64\\ 16 &
512\\ 32 & 4096\\ 64 & 32768\\
\end{tabular}
\end{ruledtabular}
\caption{Inverse temperatures needed to reach ground state properties 
for different system sizes.}
\label{tab2}
\end{table}

\begin{figure}
\begin{center}
\resizebox{!}{60mm}{\includegraphics{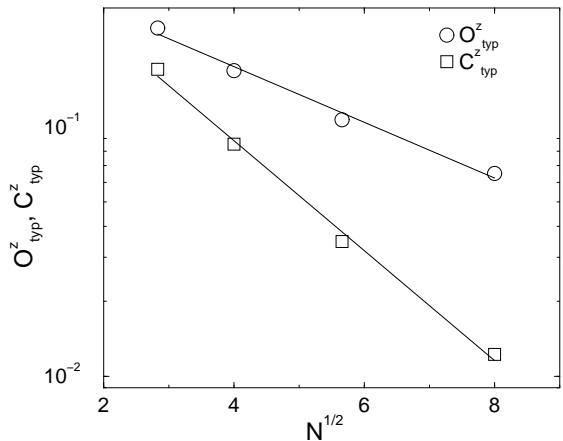}}
\end{center}
\caption{The typical spin- and string-correlation functions as a
function of the square root of the distance. The solid lines are
linear fits to the data points.}
\label{typcorrstr}
\end{figure}

In the second stage of the simulations $\mathrm{10^3}$ calibration
steps are done, followed by at least $\mathrm{2\times 10^3}$ measuring
steps. Averages are calculated over 1000 disorder
configurations. Results for average and typical correlation functions
are shown in Fig.~\ref{corrstr} and Fig.~\ref{typcorrstr}
respectively.

Straight lines are adjusted to the data points to determine the decay
exponents in Fig.~\ref{corrstr}. In the log-log plot the
string-correlation function displays very linear behavior, indicating
that the function does decay algebraically. The decay exponent is
found to be -0.378(6), in excellent agreement with the random-singlet
prediction of -0.382.  The spin-correlation function does not adjust
to a straight line as well as the string correlation. The gradient of
the curve increases with increasing system size. If a line is adjusted
it results in a decay exponent of about -1.04(1), quite different from
the predicted value of -2. The curvature of the function does
indicates that we cannot see the true decay exponent with the limited
system sizes used, and we cannot rule out the possibility that the
exponent will converge to -2 for much larger system sizes. It is,
however, remarkable that the spin-correlation function should display
such dramatic system size dependence while the string-correlation
functions has converged.

Next we consider the behavior of the typical correlation functions in
Fig.~\ref{typcorrstr}. Straight lines should result if we plot the
logarithm of the typical correlation functions versus the square root
of the systems size. The QMC data lies close to a straight line, but
there does seem to be a slight systematic curvature upwards. This can
be explained by the difficulties in measuring the typical functions. A
small percentage of the measurements are discarded since they are
negative and this will result in too large an estimate. Since more
points are dropped for the larger system sizes this should result in
an upward curvature. For the 32-site system about 0.1 \% of the points
were dropped, while about 0.6 \% of the points were dropped for the
64-site system. We believe that this explains the slight systematic
deviation from the straight line.

\begin{figure}
\begin{center}
\resizebox{!}{60mm}{\includegraphics{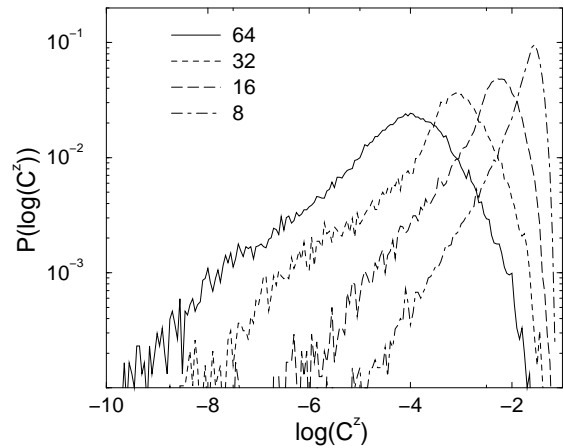}}
\end{center}
\caption{Distribution of the spin-correlation function.}
\label{distcorr}
\end{figure}

\begin{figure}
\begin{center}
\resizebox{!}{60mm}{\includegraphics{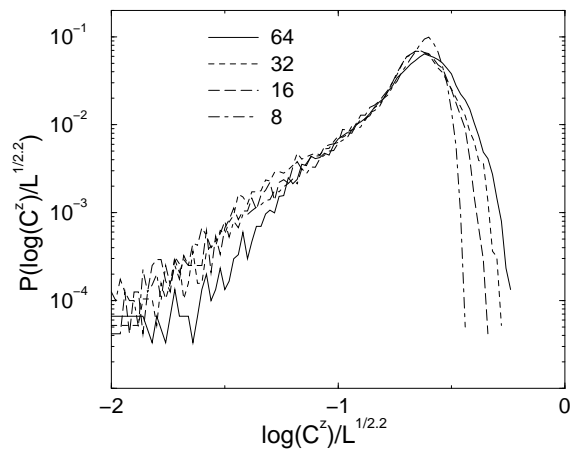}}
\end{center}
\caption{The scaled distribution function for the spin-correlation function.}
\label{scacorr}
\end{figure}

\begin{figure}
\begin{center}
\resizebox{!}{60mm}{\includegraphics{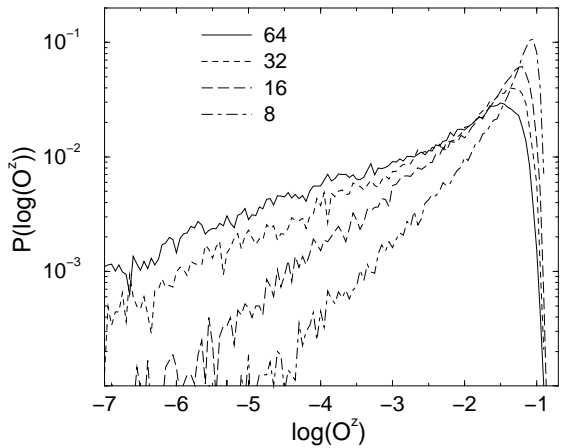}}
\end{center}
\caption{Distribution of the string-correlation function.}
\label{diststr}
\end{figure}

We also considered the distribution of the correlation functions.
According to the RG predictions the distribution of
\begin{equation}
\frac{\log\langle S_i^zS_{i+\frac{N}{2}}^z\rangle}{N^\frac{1}{\mu}}
\label{scaling}
\end{equation}
should scale to a fixed distribution with $\mu=2$. In
Fig.~\ref{distcorr} we show the distribution of the logarithm of the
spin-correlation function. The distribution becomes broader and the
peak shifts to the left with increasing $N$.  The best scaling plot
was obtained with $\mu=2.2$, shown in Fig.~\ref{scacorr}. The
agreement with the RG predictions is again very good, but we were not
able to scale the distribution of the string-correlation function. In
Fig.~\ref{diststr} the unscaled distribution is shown, and as can be
seen the distribution does broaden with increasing $N$, but much less
than for the spin correlation function.  In order to scale the
distribution so that the peaks line up, a large value of $\mu\approx
10$ is needed, but then the rest of the distribution does not scale
well.

Finally, we also studied the distribution of the local susceptibility,
\begin{equation}
\chi_{\text{loc}}=\frac{1}{N}\sum_i^N \int_0^{\beta}
\mathrm{d}\tau\langle S_i^z(0)S_i^z(\tau)\rangle,
\end{equation}
which is also predicted to scale according to Eq.~(\ref{scaling}). In
agreement with the earlier QMC study performed at higher temperatures
we also obtain the best scaling for $\mu=2.5$.

\section{Summary and Discussion}
\label{sec5}

We have done extensive QMC simulations of the antiferromagnetic spin-1
Heisenberg chain with a uniform bond distribution extending all the
way to zero bond strength. The calculations were performed at a
temperature low enough for all observables to obtain their ground
state expectation values. The decay exponent for the
string-correlation function, as well as the typical correlation
functions, are in good agreement with renormalization group
predictions for the random-singlet phase. Our results for the average
spin correlation function yields an exponent of -1.04(1) which differs
significantly from the RG prediction of -2. However, this result has
not converged, but does display some finite-size effects. The
distribution of the spin-correlation function is well described by RG
predictions, but the distribution of the string-correlation function
did not scale as predicted.

The very clear algebraic decay of the string order strongly indicates
that the system in no longer in the Haldane phase. The exponent we
obtained (-0.378(6)) is in excellent agreement with the prediction for
the random-singlet phase (-0.382), and very different from the
prediction\cite{HyPRL97} for the critical point separating the random
singlet phase from the Haldane phase (-2/3).  We therefore believe
that our results are in agreement with earlier Monte Carlo
results\cite{ToJPSJ00} and a recent RG calculation\cite{Sa02} showing
that the box distribution extending all the way to zero is enough to
drive the system to an infinite-randomness fixed point. The deviations
from the random-singlet predictions could be explained by strong
finite-size corrections. An alternative scenario is that the strong
randomness fixed point for the spin-1 chain has slightly different
properties from the spin-1/2 random-singlet fixed point.

To numerically determine the properties of the random-singlet phase
for the strongly disordered spin-1 chain is much harder than for the
spin-1/2 chain since a strong, finite disorder is needed to drive the
chain to the random-singlet phase, according to the RG predictions.
Disagreement between earlier numerical studies indicate the
difficulties.  The widest box distribution is expected to be quite
close to the transition to the Haldane state, and it is possible that
strong finite-size corrections appear in certain quantities.  It would
therefore be of great interest to study larger systems with even
stronger, power-law distribution of the bonds. This is unfortunately
not an easy task to do numerically. Density matrix renormalization
group calculations yield ground state expectation values, but the
method is most accurate for open boundary conditions for which finite
size effects are stronger. Furthermore, it is hard to keep enough
states to obtain accurate results for large systems with strong bond
disorder. Using loop quantum Monte Carlo algorithms it is possible to
reach fairly large system sizes at quite low temperatures, but to
reach temperatures low enough to accurately determine the decay of
correlation functions makes the simulations very time
consuming. Furthermore, the expectation value of the correlation
function becomes very small for large system sizes and long
simulations are needed in order to reduce the statistical errors to
the required level.

In conclusion, we believe that our results support earlier Monte Carlo
simulations\cite{ToJPSJ00} and a recent RG calculation\cite{Sa02}
indicating that the box distribution extending all the way to zero is
wide enough to reach the random-singlet phase. With our present data
we are not able to determine whether observed deviations from
predictions for the random-singlet fixed point, most notably in the
decay of the spin-correlation function, are due to strong finite-size
corrections or slightly different properties of the spin-1
random-singlet fixed point. To resolve this issue requires significant
computational efforts and we leave it for future investigations. To
obtain the present results we have implemented the concept of directed
loops for the spin-1 chain and demonstrate that autocorrelation times
are decreased by up to two orders of magnitude.

\begin{acknowledgments}
We are grateful to A. Sandvik and R. Hyman for stimulating
discussions. The work was supported by the Swedish Research Council
and the G\"oran Gustafsson foundation. P.H acknowledges support by
Ella och Georg Ehrnrooths stiftelse.
\end{acknowledgments}

\bibliography{bib} 
\end{document}